\documentclass[aps,prd,reprint,superscriptaddress,floatfix,nofootinbib]{revtex4-1}

\usepackage{dcolumn}
\usepackage{xcolor}
\usepackage{youngtab}
\usepackage{ytableau}
\usepackage{tabularray}
\usepackage{booktabs}    
\usepackage{placeins}
\usepackage{soul}
\usepackage{bm}
\definecolor{redd}{rgb}{0.8, 0.1,0.2}
\definecolor{navy}{rgb}{0.05, 0.23,0.75}
\usepackage[colorlinks=true,citecolor=red,linkcolor=blue]{hyperref}
\usepackage{chngcntr}
\hypersetup{colorlinks,	linkcolor={green!47!black},	citecolor={green!60!black},	urlcolor={green!55!black}}
\usepackage{bbm}
\usepackage{amsfonts}
\usepackage{amsmath,amssymb}
\usepackage{mathrsfs}
\usepackage{epsfig}
\usepackage{graphicx}
\usepackage{wrapfig}                 
\usepackage{url}
\usepackage[nameinlink]{cleveref}
\usepackage{float}
\usepackage{color}
\usepackage{multirow}
\usepackage{lipsum}
\usepackage{enumitem}
\newcolumntype{L}{>{\centering\arraybackslash}m{1.5cm}}

\usepackage{enumitem}
\usepackage{chngcntr}
\usepackage{tikz}

\newcommand{\be}{\begin{equation}}
\newcommand{\ee}{\end{equation}}
\newcommand{\bea}{\begin{eqnarray}}
\newcommand{\eea}{\end{eqnarray}}
\newcommand{\bc}{\begin{center}}
\newcommand{\ec}{\end{center}}

\newcommand{\orcid}[1]{\href{https://orcid.org/#1}{	\raisebox{0.5\height}{\includegraphics[height=1.25ex,width=1.25ex]{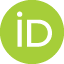}}}}

\newcommand{\commentmute}[1]{} 

\graphicspath{{Figures/}}

\begin{document}


\title{Probing Large $N_f$ Through Schemes}


\author{Alan Pinoy}
\affiliation{Dept. de Mathématique, Université Libre de Bruxelles, Boulevard du Triomphe, Bruxelles, Belgique}

\author{Shahram Vatani\,\orcid{0000-0002-1277-5829}}
\affiliation{ Centre for Cosmology, Particle Physics and Phenomenology (CP3), Université catholique de Louvain, Chemin du Cyclotron, 2 B-1348 Louvain-la-Neuve, Belgium}

\begin{abstract}
We investigate the reliability of the large $N_f$ expansion of four-dimensional gauge–fermion quantum field theories, focusing on the structure and scheme dependence of the beta function. While the existence of a nontrivial UV fixed point at leading order in $1/N_f$ suggests the possibility of asymptotic safety, the absence of higher-order terms precludes robust conclusions. We analyze the impact of renormalization scheme transformations and show that higher-order corrections inevitably introduce increasingly singular contributions. We prove that at most one renormalization scheme can preserve the dominance of the leading contribution, rendering the truncation trustworthy; in all other schemes, higher-order terms dominate and the expansion becomes unreliable. This result places strong constraints on the physical interpretation of UV fixed points in large $N_f$ theories and emphasizes the need for resummation or non-perturbative control to establish asymptotic safety.
\end{abstract}

\maketitle
The quest for ultraviolet (UV) complete quantum field theories remains a central endeavor in high-energy physics. While asymptotic freedom \cite{Gross:1973ju,Politzer:1973fx} provides one path to UV consistency, an alternative mechanism, known as asymptotic safety \cite{Weinberg:1980gg}, has attracted growing interest, particularly in the context of four-dimensional gauge-Yukawa theories \cite{Litim:2014uca}. In this framework, the theory approaches a nontrivial interacting fixed point in the UV, thereby remaining predictive at all energy scales.
These models were originally developed in the Veneziano-Witten limit, where both the number of colors and flavors are large, and exhibit perturbative control with gauge, fermionic, and scalar degrees of freedom. A natural next step is to explore whether similar UV-complete behaviors can emerge in gauge-fermion theories with fixed colors but in the many flavors limit. In particular, one aims to assess whether a large number of flavors alone can lead to asymptotic safety in gauge-fermion systems without asymptotic freedom. This question has revitalized the study of quantum field theories in the so called large  $N_f$ limit. Early investigations include large $N_f$ QED \cite{Palanques-Mestre:1983ogz} and later QCD \cite{Gracey:1996he} (see \cite{Holdom:2010qs,Gracey:2018ame} for a review). These efforts were extended to theories with multiple fermion representations \cite{Antipin:2018zdg,Cacciapaglia:2020tzd}, leading to an updated phase diagram \cite{Antipin:2017ebo}.

\vspace{0.2cm}

The appearance of asymptotic safety in these theories is tied to the existence of a nontrivial fixed point in the beta function, which can be expressed as

\begin{equation}
\beta \left( K \right) = \frac{2 K^2 }{3} \left[ 1 + \frac{F_1\left(K\right)}{N_f} + \frac{F_2\left(K\right)}{N_f^2} + \dots \right]
\label{LargeNbetafunct}
\end{equation}

\vspace{0.1cm}

where $N_f$ is the number of flavor and $K \sim \alpha N_f $ denotes the ’t Hooft coupling, rescaled version of the gauge coupling constant $\alpha$. In four-dimensional gauge-fermion systems, only the leading order contribution $F_1$ has been calculated. In the large $N_f$ limit, the vanishing of the beta function requires $F_1$ to develop a singularity, essential for the emergence of a UV fixed point. And indeed, such a singular behavior does occur at $K=3$, enabling the theory to flow toward an interacting fixed point $K^* \sim 3$ and nontrivial UV-complete regime.

\vspace{0.2cm}

It is natural to question the robustness of the fixed point found at leading order in the $1/N_f$ expansion, given the unknown impact of higher-order terms. The next-to-leading-order (NLO) contribution $F_2$ , as well as higher order terms, remains inaccessible due to the complexity of the computation, despite substantial effort to capture them, at least partially \cite{Dondi:2020qfj,Alanne:2019vuk}. An exception is the Gross–Neveu model \cite{Gracey:1993kb}, where $F_2$ has been computed and shows a singularity closer to the origin than $F_1$ , leading to a qualitatively different fixed-point structure between the $1/N_f$ and $1/N_f^2$ orders, see \cite{Antipin:2025eem} for a recent discussion.

\vspace{0.2cm}

This reflects a central limitation of truncated large $N_f$ expansions: even with access to more terms, the series lacks a clear convergence structure~\footnote{This stands in stark contrast to the more familiar perturbative expansions in a small coupling $\alpha$, where physical
quantities are approximated as power series of the form $\sum g_n \alpha^n$ . There, the general behavior $g_n \sim A^{-n} n!$ ensures that the series is meaningful up to an optimal truncation order $n \sim A /\alpha$.} , and the physical outcome can be sensitive to the shape and singularities of higher-order corrections. The NLO term, if known, could lead to three distinct scenarios:

\begin{itemize}
    \item NLO is bounded for $K \leq 3$: the fixed point persists, the NLO is negligible.
    \item NLO has a pole at $K = 3$: the outcome depends on how the leading and subleading poles interact.
    \item NLO has a pole at $K < 3$: the fixed point may be shift or disappear depending on the sign of the first singularity.
\end{itemize}

Even with a known NLO, the same uncertainty applies to NNLO and beyond. Thus, any finite-order truncation should be treated with caution; it is possible that only the full, resummed expansion captures the true physical behavior.

\vspace{0.2cm}

The objective is to demonstrate that the truncation of the large $N_f$ expansion can be reliably trusted in at most one renormalization scheme. This result follow from the structure of scheme transformations and their impact on the beta function. Even in the absence of explicit results for $F_i$ with $i \geq 2$, we will show that there exists at most one scheme in which the higher-order terms remain subleading compared to $F_1$ . In all other possible schemes, the contributions from $F_{i\geq2}$ inevitably dominate, rendering the truncation unreliable. After examining the effect of scheme transformations on the structure of the beta function, we will proceed to rigorously establish this claim.

\vspace{0.4cm}

\underline{\textbf{Scheme Transformations:}}

\vspace{0.2cm}

A scheme transformation corresponds to a redefinition of the coupling, relating two schemes via an invertible change of variables between $K$ and $\Tilde{k}$ of the form (cf. \cite{Ryttov:2012nt}):

\begin{equation}
    K= \Tilde{K} \left[ 1 + t_1 \frac{\Tilde{K}}{N_f} + t_2 \left( \frac{\Tilde{K}}{N_f}\right)^2 + t_3 \left( \frac{\Tilde{K}}{N_f}\right)^3 + \dots \right]
\label{SchemeT}
\end{equation}

The coefficients can be chosen freely as long as the transformation is invertible and they do not depend on $N_f$. We provide more details on the coupling reparametrization and the consequences of scheme redefinition within the Appendix \ref{app}.

\vspace{0.2cm}
A change of scheme induces a transformation of the beta function; however, the underlying physics, such as fixed points, must be map one-to-one between schemes. To rigorously assess the impact of higher-order corrections, we must express the $1/N_f$ expansion of the  beta function in a transformed scheme. Under the redefinition in \eqref{SchemeT}, the beta function in the new scheme, $\Tilde{\beta}$, can be formally written as:

\begin{equation}
    \tilde{\beta} \left( \tilde{K} \right) = \frac{2 \tilde{K}^2 }{3} \left[ 1 + \frac{\tilde{F}_1\left(\tilde{K}\right)}{N_f} + \frac{\tilde{F}_2\left(\tilde{K}\right)}{N_f^2} + \dots \right]
    \label{newbeta}
\end{equation}

Which is related to our original beta-function via:

\begin{equation}
     \tilde{\beta}\left( \tilde{K} \right)= \left( \frac{\mathrm{d} \tilde{K}}{\mathrm{d} K} \right) \frac{\mathrm{d} K}{\mathrm{d} \mu }  =  \left( \frac{\mathrm{d}K}{\mathrm{d}  \tilde{K} } \right)^{-1}  \beta\left(K \right)
     \label{relate}
\end{equation}

Combining \eqref{LargeNbetafunct}, \eqref{SchemeT}, \eqref{newbeta} and \eqref{relate} with the identification of powers in $1/N_f$ , one obtains a mapping between the functions $F_i$ and their counterparts in the transformed scheme, $\Tilde{F}_i$ (see Appendix \ref{app} for more details on the derivation):

\begin{align}
\tilde{F}_1\left(X\right)= & F_1\left(X\right) \label{rec1} \\[1ex]
\tilde{F}_2\left(X\right)= & F_2\left(X\right) + X^2\left( {t_1}^2-t_2 \right) + t_1 X^2 {F'}_1\left(X\right)\label{rec2} \\[1ex]
\tilde{F}_3\left(X\right)= & F_3\left(X\right) -2 X^3\left(t_1^3-2 t_2 t_1+t_3\right) \nonumber \\
&  + t_1 X^2 {F'}_2\left(X\right) \nonumber \\
& +\frac{1}{2} t_1^2 X^4 F_1''(X) + t_2 X^3 F_1'(X)  \label{rec3} \\
& +\left(t_1^2-t_2\right) X^2 F_1(X) \nonumber \\
\vdots & \nonumber \\
\tilde{F}_n\left(X\right)=& F_n\left(X\right) + \sum_{k<n} \;\; \sum_{i<n-k}  \;\; \diamond_{n,k,i} \;\; {F_k}^{\left( i\right)} \left( X\right) \label{rec} \\
\vdots & \nonumber
\end{align}

Where the coefficients $\diamond_{n,k,i}$  in the sum are polynomial in the $t_i$. We do not provide a general closed form, except for specific cases like:

\begin{align}
    \diamond_{n,1,1} &= (1-n) \;t_n \; X^n \;\;\;\; \forall \; n \geq 2
    \label{coeffn}
\end{align}

And the important relation:

\begin{equation}
    \diamond_{n,k,i}\Bigr|_{\substack{t_1, \dots , t_{n-1}=0}} = 0 \;\;
        \label{coeffnbis}
\end{equation}
For all $k \geq 2$, or $k=1$ \& $i\geq 2$.
However all coefficients can easily be computed recursively.
\vspace{0.4cm}

As expected, the known scheme invariance of $F_1$ is explicitly recovered. More importantly, a general pattern emerges in the reshuffling of the $F_i$ functions under scheme transformations. In particular, derivatives of $F_1$ appear at successively higher orders: the first derivative ${F'}_1$ enters at order $1/N_f^2$ , the second derivative ${F''}_1$ at order $1/N_f^3$ , and in general, the $(n - 1)$ -th derivative of $F_1$ contributes at order $1/N_f^n$ , see \ref{rec}. This observation plays a crucial role in our argument, as these derivatives exhibit increasingly singular behavior near the critical value $X \rightarrow 3$, dominating over lower derivatives and thus cannot be neglected. More precisely,

\begin{align}
{F_1}^{\left( n \right)}\left( X\right) \;\;\; & \underset{X \rightarrow 3}{\sim}  \;\;\; \frac{\left(-1\right)^n}{\left( X-3\right)^n}
\end{align}

Leading to:
    
\begin{align}
\frac{ {F_1}^{\left( k \right)} \left( X \right)}{ {F_1}^{\left( n \right)} \left( X \right)}  \;\;\; & \underset{X \rightarrow 3}{ \longrightarrow} \;\;\;  0 \;\;\;\; \forall k<n
\end{align}

Henceforth, a generic scheme transformation induces a series of dangerous divergences arising from successive derivatives of $F_1$. 

\vspace{0.2cm}

Consequently, the appearance of such terms after a scheme transformation poses a significant challenge to the reliability of any finite-order truncation in the large $N_f$ expansion. Armed with the above relations, our original statement can now be proven.

\vspace{0.4cm}

\underline{\textbf{Uniqueness of a Trustworthy Scheme:}}

\vspace{0.2cm}

Our previous observations naturally raise the question of whether there exists a renormalization scheme in which the truncation of the large $N_f$ expansion is indeed justified, namely, a scheme where all higher-order functions $F_i$ remain subleading compared to $F_1$. Let us assume such a scheme exists and take it as our baseline, characterized by functions $F_1, F_2, \dots$ satisfying the dominance condition

\begin{equation}
    F_i\left( X\right) \;\;  \underset{X \rightarrow 3}{ =} \;\; \mathcal{O}\left( F_1\left( X\right) \right) \;\;\; \forall i>1
    \label{domination}
\end{equation}

We now consider a transformation to a new scheme, parametrized by coefficients $t_i$, and investigate whether this dominance property can persist order by order:

\vspace{0.4cm}

$\bullet$ Starting from \eqref{rec2}, the transformed second-order function is

\begin{equation}
    \tilde{F}_2\left(X\right)=  F_2\left(X\right) + X^2\left( {t_1}^2-t_2 \right) + t_1 X^2 {F'}_1\left(X\right)
\end{equation}

Given that $\tilde{F}_1(X) = F_1(X)$, the requirement that $\tilde{F}_2$ remains subleading to $\tilde{F}_1$ near $X \to 3$ imposes the condition

\begin{equation}
\Rightarrow \;\; t_1 = 0,
\end{equation}

Since the term proportional to $F_1'(X)$, which is more singular than $F_1(X)$, would otherwise dominate.

\vspace{0.5cm}

$\bullet$ Proceeding to third order, Eq.~\eqref{rec3} yields

\begin{equation}
\tilde{F}_3(X) = F_3(X) + t_2 X^2 \left( X F_1'(X) - F_1(X) \right) - 2 t_3 X^3.
\end{equation}

Since $t_1=0$. Analogously, to preserve the subleading nature of $\tilde{F}_3$ relative to $\tilde{F}_1$, one must set

\begin{equation}
\Rightarrow \;\; t_2 = 0,
\end{equation}

As the term involving $F_1'(X)$ again introduces a more severe singularity near $X=3$.

\vspace{0.5cm}

$\bullet$ By iterating this reasoning to higher orders while making use of \eqref{coeffn} and \eqref{coeffnbis}, we find that each coefficient $t_n$ must vanish to avoid generating terms that overpower $\Tilde{F}_1$. More precisely, at each order $1/N_f^n$, the structure of the transformation enforces
\begin{equation}
\Rightarrow \;\; t_n = 0, \quad \forall n \geq 1,
\end{equation}

Which, combined with the vanishing of associated mixing coefficients, guarantees no enhancement of singular terms.

\vspace{0.4cm}

Hence, the only scheme in which the truncation of the large $N_f$ expansion can be consistently trusted, i.e., where all higher-order corrections remain subleading, is the original baseline scheme itself. This establishes the uniqueness of a “trustable” renormalization scheme in this context.

\vspace{0.2cm}

However, we have no insight into how to identify such a scheme, nor any guarantee that it exists at all. This lack of control fundamentally challenges the reliability of the large $N_f$ truncation.

\vspace{0.2cm}

Our approach extends naturally to higher-order truncations, such as in the $O(N)$ model, where the beta function also admits a $1/N$ expansion and where the functions $F_1$ and $F_2$ are explicitly known. There, as in gauge–fermion systems, $F_1$ exhibits a pole, but notably, $F_2$ develops a singularity closer to the origin and therefore dominates in the critical regime. This raises the question of whether there exists a scheme in which the second-order truncation remains reliable. Repeating the steps outlined above, one can readily show that such a scheme, if it exists, must again be unique. 

\vspace{0.2cm}

Nonetheless, the emergence of a uncontrolled poles does not preclude their potential relations to an interacting fixed point. For instance, consider a beta function of the illustrative form:

\begin{widetext}
\begin{equation}
    \beta_{\textbf{try}} \left( K \right) = \frac{2 K^2 }{3} \left[ 1 + \frac{1}{N} \frac{1}{\left( K-3 \right)} + \frac{1}{N^2} \frac{1}{2!\left( K-3 \right)^2} +\frac{1}{N^3} \frac{1}{3!\left( K-3 \right)^3} + \dots \right]
\end{equation}  
\end{widetext}

which exhibits a factorially growing tower of singular terms at each order. In this case, truncation is manifestly unreliable: higher-order contributions are increasingly divergent and dominate over lower ones. Moreover, the alternating signs of successive truncations lead to qualitatively different conclusions depending on the order at which the series is cut off. However, when summed to all orders, the series yields a well-defined resummed beta function:

\begin{equation}
   \beta_{\textbf{try}}\left( K \right) = \frac{2 K^2 }{3} \exp{\left[  \frac{1}{N\left( K-3 \right)}  \right]}\;=\;0 \;\;\; \text{at} \;\; K=3
\end{equation}

which vanishes exactly at $K = 3$. This cancellation arises from a collective resummation effect, demonstrating that an apparent divergence in the truncated expansion may still encode a genuine fixed point in the full theory. Furthermore one can show that a truncated "trustworthy" scheme cannot be reached, stressing the collective character of this illustrative beta function.

\section*{Conclusion}

In this work, we have established the structure of scheme transformations within the large $N_f$ expansion and analyzed their implications for the reliability of truncated beta functions. Our results reveal that higher-order terms inevitably introduce increasingly severe singularities which propagate through scheme transformations and dominate unless carefully canceled. Most notably, we have proven that, under very general assumptions, there exists at most a single renormalization scheme in which the truncated beta function remains under control, i.e., where higher-order contributions remain subleading. In all other schemes, the dominance of higher derivatives of the leading term renders the truncation unreliable. This highlights the stringent conditions under which asymptotic expansions in $1/N_f$ can be trusted, and point out strict limitations on the predictive power of finite-order truncations in the quest for asymptotic safety.

\begin{acknowledgments}
SV would like to thank G.Cacciapaglia for fruitful discussions. SV is supported by the Fonds de la Recherche Scientifique de Belgique (FNRS) under the IISN convention 4.4517.08 . AP is a boursier of Université Libre de Bruxelles and supported by the EoS FNRS grant ”Beyond Symplectic Geometry”. 
\end{acknowledgments}

\begin{appendix}

\section{Scheme Transformation and Resummed Beta Function Coefficients }\label{app}

We consider a scheme transformation of the coupling $\alpha \rightarrow \tilde{\alpha}$, encoded via a function $\mathcal{F}$, cf. \cite{Ryttov:2012nt}:
\begin{equation}
    \alpha = \tilde{\alpha}\,\mathcal{F}(\tilde{\alpha}) = \tilde{\alpha} + t_1\,\tilde{\alpha}^2 + t_2\,\tilde{\alpha}^3 + t_3\,\tilde{\alpha}^4 + \dots
\end{equation}

Such as the transformation is invertible. As we are working with the 't Hooft coupling $ K \equiv \alpha N_f $, we naturally define $\tilde{K} \equiv \tilde{\alpha} N_f $:

\begin{equation}
    K= \Tilde{K} \left[ 1 + t_1 \frac{\Tilde{K}}{N_f} + t_2 \left( \frac{\Tilde{K}}{N_f}\right)^2 + t_3 \left( \frac{\Tilde{K}}{N_f}\right)^3 + \dots \right]
\end{equation}

We note that certain conventions made use of gauge group factors, like the index of the representation, within the definition of $K$. Taking them into account will not change qualitatively our result.

 We can now relate the original and transformed beta functions $\beta \rightarrow \tilde{\beta}$:

\begin{align}
    \tilde{\beta}(\tilde{K}) &= \left( \frac{d\tilde{K}}{dK} \right) \frac{dK}{d\mu}
    = \left( \frac{dK}{d\tilde{K}} \right)^{-1} \beta(K) \\
    &= \left( \frac{dK}{d\tilde{K}} \right)^{-1} \frac{2 K^2}{3} \left[1 + \frac{F_1(K)}{N_f} + \frac{F_2(K)}{N_f^2} + \dots \right] \\
    &= \left( \frac{dK}{d\tilde{K}} \right)^{-1} \frac{2 \left[ \tilde{K} \mathcal{F}\left( \frac{\tilde{K}}{N_f} \right) \right]^2}{3}
    \left[1 + \frac{F_1(K)}{N_f} + \frac{F_2(K)}{N_f^2} + \dots \right]
\end{align}

We compute the derivative:
\begin{align}
    \frac{dK}{d\tilde{K}} &= \frac{d}{d\tilde{K}} \left[ \tilde{K} \mathcal{F}\left( \frac{\tilde{K}}{N_f} \right) \right] \nonumber \\
    &= \frac{d}{d\tilde{K}} \left[ \tilde{K} + \frac{t_1}{N_f} \tilde{K}^2 + \frac{t_2}{N_f^2} \tilde{K}^3 + \dots \right] \nonumber \\
    &= 1 + \frac{2t_1}{N_f} \tilde{K} + \frac{3t_2}{N_f^2} \tilde{K}^2 + \dots
\end{align}

Now the rescaled prefactor:
\begin{align}
    \frac{2 K^2}{3} &= \frac{2 \left[ \tilde{K} + \frac{t_1}{N_f} \tilde{K}^2 + \frac{t_2}{N_f^2} \tilde{K}^3 + \dots \right]^2}{3}
\end{align}

We then expand each $F_n(K)$ in terms of $\tilde{K}$:
\begin{align}
    F_n(K) &= F_n\left( \tilde{K} \mathcal{F}\left( \frac{\tilde{K}}{N_f} \right) \right) \\
    &= F_n(\tilde{K}) + \left( \frac{t_1}{N_f} \tilde{K}^2 + \frac{t_2}{N_f^2} \tilde{K}^3 + \dots \right) F_n'(\tilde{K}) \nonumber \\
    &\quad + \frac{1}{2} \left( \frac{t_1}{N_f} \tilde{K}^2 + \dots \right)^2 F_n''(\tilde{K}) + \dots
\end{align}

Combining the above relations, we find:
\begin{align*}
    \tilde{F}_1(X) &= F_1(X) \\
    \tilde{F}_2(X) &= F_2(X) + X^2(t_1^2 - t_2) + t_1 X^2 F_1'(X) \\
    \tilde{F}_3(X) &= F_3(X) - 2 X^3(t_1^3 - 2t_1 t_2 + t_3) + t_1 X^2 F_2'(X) \\
    &\quad + \frac{1}{2} t_1^2 X^4 F_1''(X) + t_2 X^3 F_1'(X) + (t_1^2 - t_2) X^2 F_1(X) \\
    \vdots &
\end{align*}

These expressions can be extended recursively. While no closed form for $\tilde{F}_n$ is captured in full generality, we observe the following structure:
\begin{equation}
    \tilde{F}_n(X) = F_n(X) + \sum_{k<n} \;\sum_{i<n-k} \diamond_{n,k,i} \cdot F_k^{(i)}(X)
\end{equation}

The coefficients $\diamond_{n,k,i}$ are polynomials in the $t_i$. A closed form can easily be captured in the specific cases:
\begin{align}
    \diamond_{n,1,n-1} &= \frac{t_1^{n-1} X^{2n-2}}{(n-1)!}, \quad \forall n \geq 2 \\
    \diamond_{n,1,1} &= (1-n) t_n X^n, \quad \forall n \geq 2
\end{align}

We also note that:

\begin{equation}
    \diamond_{n,k,i}\Bigr|_{\substack{t_1, \dots , t_{n-1}=0}} = 0 
\end{equation}
For all $k \geq 2$, or $k=1$ \& $i\geq 2$. Hence, the structure of $\tilde{F}_n$ includes terms like:
\begin{equation}
    \tilde{F}_n(X) \supset \dots + \frac{t_1^{n-1} X^{2n-2}}{(n-1)!} F_1^{(n-1)}(X) + (1-n) t_n X^n F_1'(X) + \dots
\end{equation}

In summary, given $F_k$ for $k \leq n$, one can systematically compute $\tilde{F}_n$ via recursive substitution and differentiation.

\end{appendix}

\vspace{5cm}
\bibliography{biblio}

\begin{thebibliography}{16}%
\makeatletter
\providecommand \@ifxundefined [1]{%
 \@ifx{#1\undefined}
}%
\providecommand \@ifnum [1]{%
 \ifnum #1\expandafter \@firstoftwo
 \else \expandafter \@secondoftwo
 \fi
}%
\providecommand \@ifx [1]{%
 \ifx #1\expandafter \@firstoftwo
 \else \expandafter \@secondoftwo
 \fi
}%
\providecommand \natexlab [1]{#1}%
\providecommand \enquote  [1]{``#1''}%
\providecommand \bibnamefont  [1]{#1}%
\providecommand \bibfnamefont [1]{#1}%
\providecommand \citenamefont [1]{#1}%
\providecommand \href@noop [0]{\@secondoftwo}%
\providecommand \href [0]{\begingroup \@sanitize@url \@href}%
\providecommand \@href[1]{\@@startlink{#1}\@@href}%
\providecommand \@@href[1]{\endgroup#1\@@endlink}%
\providecommand \@sanitize@url [0]{\catcode `\\12\catcode `\$12\catcode `\&12\catcode `\#12\catcode `\^12\catcode `\_12\catcode `\%12\relax}%
\providecommand \@@startlink[1]{}%
\providecommand \@@endlink[0]{}%
\providecommand \url  [0]{\begingroup\@sanitize@url \@url }%
\providecommand \@url [1]{\endgroup\@href {#1}{\urlprefix }}%
\providecommand \urlprefix  [0]{URL }%
\providecommand \Eprint [0]{\href }%
\providecommand \doibase [0]{http://dx.doi.org/}%
\providecommand \selectlanguage [0]{\@gobble}%
\providecommand \bibinfo  [0]{\@secondoftwo}%
\providecommand \bibfield  [0]{\@secondoftwo}%
\providecommand \translation [1]{[#1]}%
\providecommand \BibitemOpen [0]{}%
\providecommand \bibitemStop [0]{}%
\providecommand \bibitemNoStop [0]{.\EOS\space}%
\providecommand \EOS [0]{\spacefactor3000\relax}%
\providecommand \BibitemShut  [1]{\csname bibitem#1\endcsname}%
\let\auto@bib@innerbib\@empty
\bibitem [{\citenamefont {Gross}\ and\ \citenamefont {Wilczek}(1973)}]{Gross:1973ju}%
  \BibitemOpen
  \bibfield  {author} {\bibinfo {author} {\bibfnamefont {D.~J.}\ \bibnamefont {Gross}}\ and\ \bibinfo {author} {\bibfnamefont {F.}~\bibnamefont {Wilczek}},\ }\href {\doibase 10.1103/PhysRevD.8.3633} {\bibfield  {journal} {\bibinfo  {journal} {Phys. Rev. D}\ }\textbf {\bibinfo {volume} {8}},\ \bibinfo {pages} {3633} (\bibinfo {year} {1973})}\BibitemShut {NoStop}%
\bibitem [{\citenamefont {Politzer}(1973)}]{Politzer:1973fx}%
  \BibitemOpen
  \bibfield  {author} {\bibinfo {author} {\bibfnamefont {H.~D.}\ \bibnamefont {Politzer}},\ }\href {\doibase 10.1103/PhysRevLett.30.1346} {\bibfield  {journal} {\bibinfo  {journal} {Phys. Rev. Lett.}\ }\textbf {\bibinfo {volume} {30}},\ \bibinfo {pages} {1346} (\bibinfo {year} {1973})}\BibitemShut {NoStop}%
\bibitem [{\citenamefont {Weinberg}(1980)}]{Weinberg:1980gg}%
  \BibitemOpen
  \bibfield  {author} {\bibinfo {author} {\bibfnamefont {S.}~\bibnamefont {Weinberg}},\ }\enquote {\bibinfo {title} {{ULTRAVIOLET DIVERGENCES IN QUANTUM THEORIES OF GRAVITATION}},}\ in\ \href@noop {} {\emph {\bibinfo {booktitle} {{General Relativity}: {An Einstein Centenary Survey}}}}\ (\bibinfo {year} {1980})\ pp.\ \bibinfo {pages} {790--831}\BibitemShut {NoStop}%
\bibitem [{\citenamefont {Litim}\ and\ \citenamefont {Sannino}(2014)}]{Litim:2014uca}%
  \BibitemOpen
  \bibfield  {author} {\bibinfo {author} {\bibfnamefont {D.~F.}\ \bibnamefont {Litim}}\ and\ \bibinfo {author} {\bibfnamefont {F.}~\bibnamefont {Sannino}},\ }\href {\doibase 10.1007/JHEP12(2014)178} {\bibfield  {journal} {\bibinfo  {journal} {JHEP}\ }\textbf {\bibinfo {volume} {12}},\ \bibinfo {pages} {178} (\bibinfo {year} {2014})},\ \Eprint {http://arxiv.org/abs/1406.2337} {arXiv:1406.2337 [hep-th]} \BibitemShut {NoStop}%
\bibitem [{\citenamefont {Palanques-Mestre}\ and\ \citenamefont {Pascual}(1984)}]{Palanques-Mestre:1983ogz}%
  \BibitemOpen
  \bibfield  {author} {\bibinfo {author} {\bibfnamefont {A.}~\bibnamefont {Palanques-Mestre}}\ and\ \bibinfo {author} {\bibfnamefont {P.}~\bibnamefont {Pascual}},\ }\href {\doibase 10.1007/BF01212398} {\bibfield  {journal} {\bibinfo  {journal} {Commun. Math. Phys.}\ }\textbf {\bibinfo {volume} {95}},\ \bibinfo {pages} {277} (\bibinfo {year} {1984})}\BibitemShut {NoStop}%
\bibitem [{\citenamefont {Gracey}(1996)}]{Gracey:1996he}%
  \BibitemOpen
  \bibfield  {author} {\bibinfo {author} {\bibfnamefont {J.~A.}\ \bibnamefont {Gracey}},\ }\href {\doibase 10.1016/0370-2693(96)00105-0} {\bibfield  {journal} {\bibinfo  {journal} {Phys. Lett. B}\ }\textbf {\bibinfo {volume} {373}},\ \bibinfo {pages} {178} (\bibinfo {year} {1996})},\ \Eprint {http://arxiv.org/abs/hep-ph/9602214} {arXiv:hep-ph/9602214} \BibitemShut {NoStop}%
\bibitem [{\citenamefont {Holdom}(2011)}]{Holdom:2010qs}%
  \BibitemOpen
  \bibfield  {author} {\bibinfo {author} {\bibfnamefont {B.}~\bibnamefont {Holdom}},\ }\href {\doibase 10.1016/j.physletb.2010.09.037} {\bibfield  {journal} {\bibinfo  {journal} {Phys. Lett. B}\ }\textbf {\bibinfo {volume} {694}},\ \bibinfo {pages} {74} (\bibinfo {year} {2011})},\ \Eprint {http://arxiv.org/abs/1006.2119} {arXiv:1006.2119 [hep-ph]} \BibitemShut {NoStop}%
\bibitem [{\citenamefont {Gracey}(2019)}]{Gracey:2018ame}%
  \BibitemOpen
  \bibfield  {author} {\bibinfo {author} {\bibfnamefont {J.~A.}\ \bibnamefont {Gracey}},\ }\href {\doibase 10.1142/S0217751X18300326} {\bibfield  {journal} {\bibinfo  {journal} {Int. J. Mod. Phys. A}\ }\textbf {\bibinfo {volume} {33}},\ \bibinfo {pages} {1830032} (\bibinfo {year} {2019})},\ \Eprint {http://arxiv.org/abs/1812.05368} {arXiv:1812.05368 [hep-th]} \BibitemShut {NoStop}%
\bibitem [{\citenamefont {Antipin}\ \emph {et~al.}(2018)\citenamefont {Antipin}, \citenamefont {Dondi}, \citenamefont {Sannino}, \citenamefont {Thomsen},\ and\ \citenamefont {Wang}}]{Antipin:2018zdg}%
  \BibitemOpen
  \bibfield  {author} {\bibinfo {author} {\bibfnamefont {O.}~\bibnamefont {Antipin}}, \bibinfo {author} {\bibfnamefont {N.~A.}\ \bibnamefont {Dondi}}, \bibinfo {author} {\bibfnamefont {F.}~\bibnamefont {Sannino}}, \bibinfo {author} {\bibfnamefont {A.~E.}\ \bibnamefont {Thomsen}}, \ and\ \bibinfo {author} {\bibfnamefont {Z.-W.}\ \bibnamefont {Wang}},\ }\href {\doibase 10.1103/PhysRevD.98.016003} {\bibfield  {journal} {\bibinfo  {journal} {Phys. Rev. D}\ }\textbf {\bibinfo {volume} {98}},\ \bibinfo {pages} {016003} (\bibinfo {year} {2018})},\ \Eprint {http://arxiv.org/abs/1803.09770} {arXiv:1803.09770 [hep-ph]} \BibitemShut {NoStop}%
\bibitem [{\citenamefont {Cacciapaglia}\ and\ \citenamefont {Vatani}(2021)}]{Cacciapaglia:2020tzd}%
  \BibitemOpen
  \bibfield  {author} {\bibinfo {author} {\bibfnamefont {G.}~\bibnamefont {Cacciapaglia}}\ and\ \bibinfo {author} {\bibfnamefont {S.}~\bibnamefont {Vatani}},\ }\href {\doibase 10.1140/epjc/s10052-021-09257-8} {\bibfield  {journal} {\bibinfo  {journal} {Eur. Phys. J. C}\ }\textbf {\bibinfo {volume} {81}},\ \bibinfo {pages} {476} (\bibinfo {year} {2021})},\ \Eprint {http://arxiv.org/abs/2005.07540} {arXiv:2005.07540 [hep-ph]} \BibitemShut {NoStop}%
\bibitem [{\citenamefont {Antipin}\ and\ \citenamefont {Sannino}(2018)}]{Antipin:2017ebo}%
  \BibitemOpen
  \bibfield  {author} {\bibinfo {author} {\bibfnamefont {O.}~\bibnamefont {Antipin}}\ and\ \bibinfo {author} {\bibfnamefont {F.}~\bibnamefont {Sannino}},\ }\href {\doibase 10.1103/PhysRevD.97.116007} {\bibfield  {journal} {\bibinfo  {journal} {Phys. Rev. D}\ }\textbf {\bibinfo {volume} {97}},\ \bibinfo {pages} {116007} (\bibinfo {year} {2018})},\ \Eprint {http://arxiv.org/abs/1709.02354} {arXiv:1709.02354 [hep-ph]} \BibitemShut {NoStop}%
\bibitem [{\citenamefont {Dondi}\ \emph {et~al.}(2020)\citenamefont {Dondi}, \citenamefont {Dunne}, \citenamefont {Reichert},\ and\ \citenamefont {Sannino}}]{Dondi:2020qfj}%
  \BibitemOpen
  \bibfield  {author} {\bibinfo {author} {\bibfnamefont {N.~A.}\ \bibnamefont {Dondi}}, \bibinfo {author} {\bibfnamefont {G.~V.}\ \bibnamefont {Dunne}}, \bibinfo {author} {\bibfnamefont {M.}~\bibnamefont {Reichert}}, \ and\ \bibinfo {author} {\bibfnamefont {F.}~\bibnamefont {Sannino}},\ }\href {\doibase 10.1103/PhysRevD.102.035005} {\bibfield  {journal} {\bibinfo  {journal} {Phys. Rev. D}\ }\textbf {\bibinfo {volume} {102}},\ \bibinfo {pages} {035005} (\bibinfo {year} {2020})},\ \Eprint {http://arxiv.org/abs/2003.08397} {arXiv:2003.08397 [hep-th]} \BibitemShut {NoStop}%
\bibitem [{\citenamefont {Alanne}\ \emph {et~al.}(2019)\citenamefont {Alanne}, \citenamefont {Blasi},\ and\ \citenamefont {Dondi}}]{Alanne:2019vuk}%
  \BibitemOpen
  \bibfield  {author} {\bibinfo {author} {\bibfnamefont {T.}~\bibnamefont {Alanne}}, \bibinfo {author} {\bibfnamefont {S.}~\bibnamefont {Blasi}}, \ and\ \bibinfo {author} {\bibfnamefont {N.~A.}\ \bibnamefont {Dondi}},\ }\href {\doibase 10.1103/PhysRevLett.123.131602} {\bibfield  {journal} {\bibinfo  {journal} {Phys. Rev. Lett.}\ }\textbf {\bibinfo {volume} {123}},\ \bibinfo {pages} {131602} (\bibinfo {year} {2019})},\ \Eprint {http://arxiv.org/abs/1905.08709} {arXiv:1905.08709 [hep-th]} \BibitemShut {NoStop}%
\bibitem [{\citenamefont {Gracey}(1994)}]{Gracey:1993kb}%
  \BibitemOpen
  \bibfield  {author} {\bibinfo {author} {\bibfnamefont {J.~A.}\ \bibnamefont {Gracey}},\ }\href {\doibase 10.1142/S0217751X94000285} {\bibfield  {journal} {\bibinfo  {journal} {Int. J. Mod. Phys. A}\ }\textbf {\bibinfo {volume} {9}},\ \bibinfo {pages} {567} (\bibinfo {year} {1994})},\ \Eprint {http://arxiv.org/abs/hep-th/9306106} {arXiv:hep-th/9306106} \BibitemShut {NoStop}%
\bibitem [{\citenamefont {Antipin}\ \emph {et~al.}(2025)\citenamefont {Antipin}, \citenamefont {Pinoy}, \citenamefont {Sannino},\ and\ \citenamefont {Vatani}}]{Antipin:2025eem}%
  \BibitemOpen
  \bibfield  {author} {\bibinfo {author} {\bibfnamefont {O.}~\bibnamefont {Antipin}}, \bibinfo {author} {\bibfnamefont {A.}~\bibnamefont {Pinoy}}, \bibinfo {author} {\bibfnamefont {F.}~\bibnamefont {Sannino}}, \ and\ \bibinfo {author} {\bibfnamefont {S.}~\bibnamefont {Vatani}},\ }\href@noop {} {\  (\bibinfo {year} {2025})},\ \Eprint {http://arxiv.org/abs/2504.05988} {arXiv:2504.05988 [hep-ph]} \BibitemShut {NoStop}%
\bibitem [{\citenamefont {Ryttov}\ and\ \citenamefont {Shrock}(2012)}]{Ryttov:2012nt}%
  \BibitemOpen
  \bibfield  {author} {\bibinfo {author} {\bibfnamefont {T.~A.}\ \bibnamefont {Ryttov}}\ and\ \bibinfo {author} {\bibfnamefont {R.}~\bibnamefont {Shrock}},\ }\href {\doibase 10.1103/PhysRevD.86.085005} {\bibfield  {journal} {\bibinfo  {journal} {Phys. Rev. D}\ }\textbf {\bibinfo {volume} {86}},\ \bibinfo {pages} {085005} (\bibinfo {year} {2012})},\ \Eprint {http://arxiv.org/abs/1206.6895} {arXiv:1206.6895 [hep-th]} \BibitemShut {NoStop}%
\end{thebibliography}%

\end{document}